# THE JAHN-TELLER EFECT: A PERMANENT PRESENCE IN THE FRONTIERS OF SCIENCE

R. Englman[a,b] and A. Yahalom[a]
[a]*College of Judea and Samaria, Ariel 44284, Israel*
[b]*Department of Physics and Applied Mathematics,
Soreq NRC, Yavne 81800, Israel*

**Abstract:** In 1937 the Jahn-Teller (JT) effect addressed the instability (potential or actual) of non-linear symmetric molecules with degenerate orbital electronic states. In view of the large variety of JT activity that has taken place since then, we might broaden our perspective to look at works whose subjects fall under the more general heading of "Strong interactions between two dissimilar systems" (where one system is usually bosonic and the other fermionic). In these intervening years we find several highly important works in Physics and Chemistry that come under this heading and were either connected with, or arose from, JT systems, problems and approaches. Apart from high temperature superconductors, we mention Yang-Mills gauge-forces, symmetry breaking (in elementary particles), conical intersections in molecular potential surfaces, surface crossings between them in chemical reactions, entanglements in the quantum theory of measurements and Berry phases. We elaborate on the last two topics. We show first that the slow evolution of a $T\otimes\epsilon$ coupling from the weak to strong regime can model the quantum mechanical three-state measurement situation, when the positions of the nuclei acts as the measuring device. We then employ recently derived integral relations between component moduli and phases in a time dependent wave-function to demonstrate the equivalence between the state-reduction and the phase decoherence interpretations of the measurement process.

## 1. INTRODUCTION

By achieving a working separation between interacting particles of different kinds, the Born-Oppenheimer (BO) scheme devised in 1927 has been hailed as one of the greatest advances in theoretical physics. However, when one looks at theoretical physics in the year of 2000 (and several years preceding that) one notes a trend in the opposite direction, towards unification of all forces and masses at a basic level, suggesting that the observed differences in these arise (or have arisen in the distant past) as a sort of accident, by the breaking of a fundamental symmetry, which has yet to be fully revealed. The BO scheme has no place in this primordial state. The JT effect fits in very snugly in this development, first by the simplistic fact that it is effective when the BO scheme is not, but also since one of its achievements is to show how a non-symmetric situation originates in a symmetric source.

It is thus no wonder that so many developments of modern theoretical physics bear some relation to the JT effect. Less evident, but still indisputable, are its traces in experimental physics and in chemical processes. We shall recall some instances of these, highlighting the affinities, but without going into great details (which does not seem possible in the scope of a single article). In addition to noting the affinities, we shall indicate the difference in each topic between a conventional JT approach and the actual one.

## 2. STRONGLY COUPLED FERMIONS AND BOSONS
### 2.1. High Temperature Superconductivity

The motivational role of the JT effect is clearly documented in the early papers of Bednorz and Muller [1,2] and this appears to account for the significant resurgence of interest in the JT effect, immediately following the discovery of high temperature superconductivity [3]. In the course of time, theories of the phenomenon have proliferated and moved away from the original concept to larger or lesser extents, depending on the nature of the theories.

### 2.2. Entanglement in Quantum Measurements [4]

The measurement process is represented as taking place in three stages:
I. Preparation of the "small" microscopic system in a superposition state.
II. By bringing the system in interaction with the "large" measuring device, creation of an entangled device-system superposition state.
III. Observation of the device state and deduction of the system state.

In the three stages the wave function is in the following (not normalized) form:
I. $\Psi$= |device state> ( |x>+|y>+|z>)
II. $\Psi$= |device state1> |x>+|d.s.2> |y> + |d.s.3> |z>
III. $\Psi$ = |device state1>$_{observed}$ :
    (future) system state: |x>

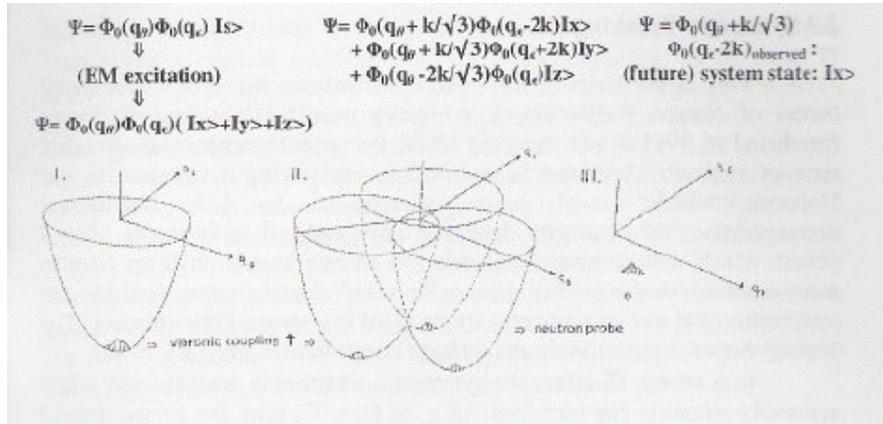

*Figure1.* A Jahn-Teller quantum measurement set-up, explained in the text.

The process can be conveniently illustrated through a $T\otimes\epsilon$ JT system in which the "small" system is the electronic triplet (labeled x,y,z) and the "large" system is the $\epsilon$-type motion of the surrounding nuclei.

To reach stage I, an excitation is made from (say) the ground vibronic level of an s-type electronic state to the triplet, using three coherent electric fields polarized along three orthogonal directions. The JT coupling within the triplet is assumed to be negligible at this stage. (The states are shown in the figure.) The coupling is then slowly turned on, eventually to reach the strong JT limit. The adiabatic development of the system keeps it in the ground state (shown at stage II in the figure). The three nuclear states in the three positions are practically orthogonal (which property characterizes a macroscopic device) and their probing yields a unique answer. This determines the "measured" electronic state, (stage III in figure) which will be one of the three orthogonal states, each occurring, according to the rules of quantum mechanics, with a probability of 1/3. We emphasize that the high symmetry is to be strictly maintained in the procedure.

The advantage of the foregoing JT illustration is that it represents a combination of device-system whose details are quite well known, in contrast to other measuring devices, which are usually very complex. Because of this, it might be possible to model both the entanglement process (that leading from stage I to II) and the wave-packet collapse (between II and III) in a precise way. The mechanism of the collapse is acknowledged to be the most problematic issue in quantum mechanics. Current explanations are: the Copenhagen- von Neumann (state-reduction) interpretation involving the observer, the "many world interpretation" (where every measurement outcome has its own wave-function), the interaction with a stochastic environment (which destroys the phase coherence between states in the superposition) and some hidden variable theories. All are controversial.

## 2.3 Symmetry Breaking [5]

This is the basis of differentiation between the four fundamental forces of Nature. Reduction of symmetry entails Higgs bosonic forces (predicted in 1961-4 and expected since, but as yet unobserved). Another type of symmetry breaking is involved in introducing a structure in our Universe, which is, of course, prerequisite for Life. Traditional demonstrations of symmetry breaking have utilized a vertically placed pencil, which will ultimately fall down to an asymmetric situation (due to some random exterior forces). A more "modern" demonstration would be the localization due to low symmetry strain fields in a strong $E \otimes \epsilon$ situation. The analogy between these fields and a Higgs boson field is straightforward:

In a strong JT effect the symmetric situation is unstable and a low symmetry situation can take over. (E.g. in $O_h$ a $T_{2g}$ state can be the ground state). From an observed, permanent reduction of symmetry, one deduces the presence of some low symmetry field. (This field is "boson-type", e.g., due to other atoms in the neighborhood.) The nature of this low symmetry field is given (more or less) by the nature of the distortion (They are shown in the back inside cover of a book [6].) In fundamental particle physics also, the symmetry breaking is necessarily associated with an additional force: the Higgs boson. Two quotations from [5] are of value:

*"Broken symmetry is always associated with a degeneracy". "There must be external perturbations greater than the energy differences between rotational levels. For infinite sized systems there are zero matrix elements between different vacua."* (This means that there is no tunneling)

However, there are differences between JTE and particle physics: The boson is coupled also to another field (not present in the JTE). The second ("upper") fermionic state (which is so obvious in the JT picture) is not directly put in: its presence is only felt through the assumption of a "Mexican-hat" potential, function of the bosonic vector field $\boldsymbol{\phi}$ (<u>not</u> an angle), having the form

$$V(\phi) = -a\,\boldsymbol{\phi}^2 + b\boldsymbol{\phi}^4 \tag{1}$$

However if we look for a plausible origin of the (phenomenological) negative quadratic term, then this can come, by second order perturbation, from upper fermionic states. So, there are other fermionic states!

## 2.4. Yang-Mills Gauge Forces [7]

These achieve the unification of three basic forces (excluding gravitation). Originally [8], they were applied to the proton-neutron doublet

$$\Psi = \begin{pmatrix} \psi_p \\ \psi_n \end{pmatrix} \qquad (2)$$

but they are applicable also for all non-abelian fields (fields whose Hamiltonians contain matrices that are non-commuting, as in most JT systems [6,9]). By introducing local gauge variations (i.e., a gauge which is a <u>general</u> function of the position), the requirement on the symmetry of the Lagrangean (or Hamiltonian) entails the presence of a boson field (a meson). Apart from the notorious phase factor (that we shall discuss in the section on Berry phase), the corresponding gauge force is absent in the JT effect, in which there is a restriction to low dimensionality (between 2– 5) of the Hilbert space. This does not allow a general space variation.

[Interestingly, in contrast to the fundamental importance of the Yang-Mills gauge fields, solutions of the Yang-Mills equations are not available. (They are one of the items in the Clay-Mathematical Institute prize problems in 2000.) It may be that the JT community can make some profit here.]

### 2.4. Curve Crossing Treatments in Chemical Kinetics

Several subjects of considerable basic and practical importance were studied. These relate to JT systems, in particular to the $(A+A+ ..) \otimes (\alpha+\alpha +…)$ pseudo-JT type, (where the A's are, possibly unstable, electronic states and $\alpha$ represents a reaction coordinate R.) The problems and solutions are time dependent [10]. The traditional (frequently, semi-classical) treatments are associated with the names Landau-Zener-Stueckelberg. However, recently they were put into explicit, algebraic forms so that a comparison with the JT formalism can be readily made. Following Child's treatment, other workers studied curve-crossing problems, e.g., [11]. In these, the system possessing an energy E slides down on a potential curve (actually a surface) along a reaction coordinate R, until it approaches a crossing point $R_c$ to another curve. The probability of curve crossing is of interest and this is given in terms of rather complicated expressions. However, these depend essentially on two parameters only, a and b. We shall now interpret these parameters in the (to us) familiar JT coupling scheme:

After separating off the "average" slope ($k_0R$), one obtains near the crossing point $R_c$ a Hamiltonian matrix and the eigenvalues given by:

$$H = \begin{pmatrix} -k(R-R_c) & \Delta E/2 \\ \Delta E/2 & k(R-R_c) \end{pmatrix} - k_0R \qquad (3)$$

$$-k_0R \pm \sqrt{(\Delta E/2)^2 + [k(R-R_c)]^2} \qquad (4)$$

Fundamental parameters in the analytic expressions for probabilities of crossing between potentials are:

$$a^2 = k(k_0^2-k^2)/ \Delta E^3 \qquad (5)$$
$$b^2 = Ek/\Delta E \sqrt{(k_0^2-k^2)} \qquad (6)$$

From these we note that only pseudo-JT parameters, near the crossing point, and the kinetic energy E of reactants appear in the expressions.

## 2.5. Electron Transfer Reactions

A formally related topic is the crossing between free-energy surfaces, resulting in electron transfer in solutions. The rate of reaction varies markedly with the positioning of the crossing point with respect to the minimum on the initial surface, with the fastest reaction rate being achieved when the crossing coincides with the minimum. The explanation is due to R.A. Marcus [12], but quantal formulations in terms of the above pseudo-JT scheme $(A+A+ ..)\otimes(\alpha+\alpha +…)$ were given by others [13]. In these, the $\alpha$'s are solvent-particle coordinates, which in the thermodynamic limit pass into a continuum. Frequently, it is possible to simplify the problem by use of an effective coordinate, familiar in multi-mode JT problems [6,9].

## 2.6. The Berry (or Topological) phase [14].

For this we shall adopt a definition from the book by Chancey and the late Mary O'Brien [15]:*"The phase that can be acquired by a state moving adiabatically (slowly) around a closed path in the parameter space of the system"*. There is a further, somewhat more general phase, that appears in any cyclic motion (not necessarily slow) in the Hilbert space, which is called the Aharonov-Anandan phase.

It has been shown, in [16] and later by others, that this phase can be realized in an $E\otimes\epsilon$ Jahn Teller system. We now explore the further, almost inevitable question: whether one <u>needs</u> "another" state to obtain a phase upon cycling? Views differ on this. {Berry's ground-laying paper [14] discusses degeneracies, but his expression for the phase change makes no reference to a partner state, and neither do some other works, especially those exhibiting Berry phases in extended systems, e.g. [17]. From the result that we now present it emerges that the partner state can be disregarded, if one looks only at the result after a full revolution. However, from the situations at interim stages, the existence of a partner state is evident.}

Our approach rests on a pair of "Reciprocal relations", that show that if the phase undergoes change, so does the amplitude and *vice versa* [18-19]. The $E\otimes\epsilon$ JT problem serves as a convenient illustration. Its solution is well known, namely,

$\Psi(\varphi) = \cos(\varphi/2) |1\rangle + \sin(\varphi/2) |2\rangle$ (7)

where $|1\rangle$ and $|2\rangle$ are the partner electronic kets, $\varphi$ is the angular coordinate in the plane of the $\epsilon$-modes. The chosen solution is such that at $\varphi=0$ the system is in the $|1\rangle$ state. The relevant reciprocal relation is between the modulus of the amplitude and the relative phase ("*arg*") of the components:

$ln|\cos(\varphi/2)| = -(1/\pi)P\int d\varphi' arg[(\cos(\varphi'/2)]/(\varphi'-\varphi)$ (8)

where P stands for the principal value of the singular integral and the range of integration is over $(-\infty, \infty)$. A general proof was also given for a time dependent superposition of the (t=0) energy eigenstates, having the form.

$\Psi(t)=a(t)|1\rangle + b(t)|2\rangle + c(t)|3\rangle + \ldots$ (9)

The following relations hold for each amplitude [e.g., a(t)] separately, provided these satisfy certain conditions of analyticity in the complex t-plane (It was shown that those conditions hold for the ground state of a <u>nearly</u> adiabatically developing state and for a range of coherent wave-packets.):

$ln|a(t)| = -(1/\pi)P\int dt' arg[a(t')]/(t'-t)$ (10)

$arg\ a(t) = (1/\pi)P\int dt'\ ln|a(t')|/(t'-t)$ (11)

(The relations resemble Kramers-Kronig dispersion relations, which however operate in the frequency domain and are due to causality.) We therefore conclude that the wave function amplitude necessarily changes when the phase does. Therefore, to conserve normalization, there must be (at least) one other component. We also note that by eq. (11), "*arg*(t)", the relative phase, must be an observable quantity, since the modulus in the integrand is such for all values of t. A related result was recently proven in [20], namely, that upon circling around a conical intersection, there is an even number of states that change signs, and an even or odd number of states that do not change sign.

### 2.6.1 Equivalence of Collapse Mechanisms

We now hypothesize that the reciprocal relations (10)- (11) can be applied to the measurement situation. Let $t=T_M$ be the time of state reduction, when one of the states, say $|1\rangle$, is observed. Then, all coefficients b(t), c(t),.. , *except* a(t) (say), undergo jumps from a finite value to a very small value (close to zero), so that, e.g., $ln|b(t)|$ changes instantaneously by a large negative quantity. For reasons that will be soon clear, we shall denote this quantity by $-2N\pi$, where N is a large integer. Thus

$ln|b(T_M+0)| - ln|b(T_M-0)| = -2N\pi$ (12)

We shall now obtain an expression that (*i*) has this form and (*ii*) is the real part of a function *ln* b(t) that has the required analytic properties (it is regular in the lower half of the t-plane). We shall then derive the imaginary part of *ln*b(t), namely the phase, and shall show that the two (real and imaginary) parts of the logarithm consistently reconstruct the dual description of a measurement situation. Now it has been shown in [18] that the following functions are Hilbert transforms (meaning that they satisfy the reciprocal relations in (10)-(11)) for functions that are periodic with a period of $2T_M$

$$f(t) = \log[\cos^2(\pi t/2T_M)] \tag{13}$$
$$g(t) = -2\pi \, St(t, 2T_M) + \pi t/T_M \tag{14}$$

where St(t,T) is the step function (being zero for t<T and increasing by unity at positive increments of $T_M$ in t) and g(t) itself is the "Saw-Tooth" function. This is equivalent to (9), but to better see how the Hilbert transforms arise, we note that

$$\lim_{\varepsilon \to 0+} i\{0.5 \, ln[1+(1-\varepsilon)\exp(-i\pi t/T_M)]\} \to g(t)+if(t) \tag{15}$$

and that the real and imaginary parts of a function having certain analytic properties (namely, regularity in the lower half of the complex t-plane and the vanishing on a large semicircle in that half) are Hilbert transforms [21]. The left hand side of (15) satisfies these conditions.

We now take for the phase of the coefficient b(t) the quantity
$$arg \, b(t) = Nf(t) \tag{16}$$
and for the logarithm of the modulus
$$ln \, |b(t)| = Nf(t) \tag{17}$$

Trivially, these two [being mere multiples of (13) and (14)]) have also the Hilbert transform properties. Figures 2(a) and 2(b) show the singular behaviors of the log Modulus [$ln \, |b(t)|$] and of the phase [$arg \, b(t)$] (plotted with N=1). We see that the phase factor exp[i*arg*b(t)] oscillates very strongly when t is near $T_M$. [In (13) the argument of the logarithm is then zero, since $\cos(\pi/2)=0$.] Therefore the mean phase factor $<\exp[i arg b(T_M)]>$ (being an average over collapse-times $T_M$) tends to zero. We have thus shown that the analytic character of the time dependent wave-function [expressed through the reciprocal relations (11)-(12)] forces an equivalence between the two interpretations of the collapse noted in section **2.2**, namely, the vanishing of all but one the amplitude and the decoherence of relative phases.

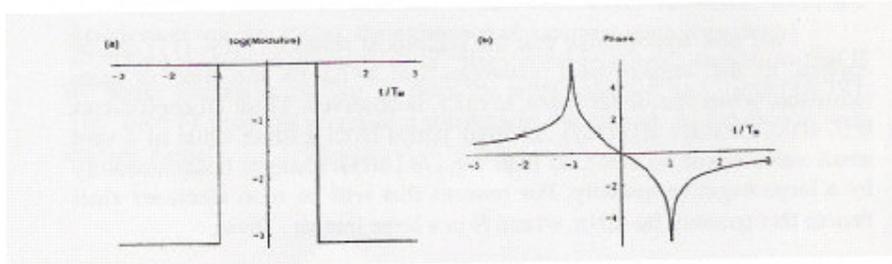

*Figure 2*. The *s*ingular behaviors of the moduli (a) and of the phase (b) at the instant of the collapse (t=$T_M$).

Is the converse also true, namely that "dephasing" (meaning, *any* sort of dephasing) entails the vanishing of amplitude? We believe (though we do not have a formal proof for this) that at least an instantaneous *jump* (of some magnitude) is an inevitable consequence of dephasing. The reason is that, probably, functions of the form given in (15) are the only ones that (upon passing to the limit $\varepsilon \to 0+$) give a divergent phase at $t=T_M$ and are solutions of a time dependent Schrodinger equation. (It is assumed that being such a solution requires certain analytic properties to be satisfied by it.) However, the Hilbert transform of the phase f(t) in (13), or Nf(t) in (16), is a "Saw –Tooth" function, and this manifests at $t=T_M$ an abrupt change in the modulus.

Admittedly, the above derivation needs some filling in of details, of both mathematical and physical kinds. In particular, one needs to describe events that occur after $2T_M$, where the formalism [which in its present form is periodic, as apparent from (13) ] describes further negative steps.

## 3. Conclusion

We have focussed on a number of highly influential works in Physics and Chemistry, which are either intrinsically or circumstantially related to the JT effect. Not in all the cases is the affinity transparent and, by suggesting some reformulation, we have tried to narrow some gaps in the correspondence. It is also admitted that the list could be further enlarged (cf. [22]). The main moral from the wide span of examples is that we are beginning to appreciate that the JT effect, which started out as a molecular phenomenon, is anchored in some deep truth of Quantum Theory.

### *REFERENCES*

[1] J.G. Bednorz and K.A. Muller, Z. Phys. B **64** (1986) 189-193 and in "Nobel Lectures: Physics" (World Scientific, Singapore, 1993), G. Ekspong (editor), pp. 424-444: "*The guiding idea in developing the concept* [that $T_c$ would increase if the electron-phonon interaction and the carrier concentration n($E_F$) would increase] *was influenced by the*

*Jahn-Teller (JT) polaron model…. We started the search for high-$T_c$ superconductivity in the late summer 1983 with the La-Ni-O system. LaNiO$_3$ is a metallic conductor with the transfer*
*energy of the JT-$e_g$ electron larger than the JT stabilization energy."*
 [2] K.A. Muller and J.G. Bednorz, Science, **237** (1987) 1133-1139. *"A mechanism for polaron formation is the Jahn Teller effect…Isolated $Fe^{4+}$ ,$Ni^{3+}$, $Cu^{2+}$ in octahedral oxygen environment show strong Jahn Teller effects…*

*Here we report on the synthesis and electrical measurements of compounds within the Ba-La-Cu-O system. This system exhibits a number of oxygen-deficient places with mixed-valent copper constituents, i.e. with itinerant electronic states between the non-JT $Cu^{3+}$ and the JT $Cu^{2+}$ ions…"*

[3]  Based on subject listings in Physics Abstracts , Years 1980 –1998. The number of JT listings has been divided by the total number of listings in the year. A significant increase in the relative number of JT listings starts at 1986 and persists for about 5 years.
[4] B. D.Espagnat, "Conceptual Foundations of Quantum Mechanics" (Benjamin, Menlo Park, 1997); P. Busch, P. J. Lahti and P. Mittelstaedt,"The Quantum Theory of Measurements" (Springer-Verlag, Berlin, 1991) (Contains extensive references.)
[5] S. Weinberg. "Quantum Theory of Fields" (University Press,. Cambridge, 1995) Chapter 11
[6] R. Englman, "The  Jahn-Teller Effect in Molecules and Crystals" (Wiley-Interscience, London, 1972)
[7] J. Leite Lopes, "Gauge Field Theories"  (Pergamon, Oxford, 1981)
[8] C.N. Yang and R.L. Mills, Phys. Rev. **96** (1954) 191-195
[9] I.B. Bersuker and V.Z. Polinger, "Vibronic Interactions in Molecules and Crystals" (Springer-Verlag, Berlin 1989) (section 3.2.1)
[10] M.S. Child, "Semiclassical Mechanics with Molecular Applications" (Clarendon Press. Oxford, 1991)
[11] Ch. Zhu, H. Nakamura, N. Re and V. Aquilanti, J. Chem. Phys. **97** (1992) 1892-1904; Ch. Zhu and H. Nakamura, Chem. Phys. Letters, **258** ( 1996) 342-347.
[12] R. A. Marcus, J. Chem. Phys. **43** (1965) 679-701 (and references therein)
[13] N.R. Kestner, J. Logan and J. Jortner, J. Phys. Chem. **78** (1974) 2148-2166
[14] M.V. Berry, Proc. Roy. Soc. A **392** (1984) 45-57
[15] C.C. Chancey and M.C. M. O'Brien, "The Jahn-Teller Effect in $C_{60}$ and Other Icosahedral Complexes" (University Press, Princeton,1997)
[16] F.S. Ham, Phys. Rev. Letters **58** (1987) 725-728
[17] R. Resta, Phys. Rev Letters **80** (1998) 1800-1803.
[18] R. Englman, A. Yahalom and  M. Baer, J. Chem. Phys. **109** (1998) 6550-5; Phys. Letters A **251** (1999) 223-8; Europ. Phys. J. D **8** (2000) 1-7
[19] R. Englman and A. Yahalom, Phys. Rev. A **60** (2000) 1802-1810.
[20] M. Baer, Chem. Phys. Letters **322** (2000) 520-526.
[21] E.C. Titchmarsh, "Introduction to the Theory of Fourier Integrals" , (Clarendon Press, Oxford, 1948)
*[22] W.H. Louisell, "Quantum Statistical Properties of Radiation" (Wiley, New York, 1990) (Especially, the sections on spin-phonon coupling and the Jaynes-Cummings model.)*